\documentclass[%
 reprint,
 amsmath,amssymb,
 aps,
 pre,
]{revtex4-1}

\usepackage{graphicx}
\usepackage{dcolumn}
\usepackage{bm}

\begin{document}


\title{Linear stability analysis of self-gravitating granular gas}

\author{Tomohiro Tanogami}
\affiliation{%
Department of physics, Kyoto University, Kyoto 606-8502, Japan
}%





\begin{abstract}
The linear stability of granular gas that reflects the contribution of self-gravitational force of mass density perturbations is investigated in order to clarify the condition of competition between clustering instability and Jeans instability.
It is found that the condition depends on three parameters: the mass density $\rho_0$, the collision rate $\omega_0$, and the rate of energy loss per collision $\epsilon$.
When $\sqrt{G\rho_0}\ll\epsilon\omega_0$, clustering instability dominates, while when $\epsilon\omega_0\ll\sqrt{G\rho_0}$, Jeans instability dominates.
These instabilities are characterized by the decrease and increase, respectively, of the temperature.  
\end{abstract}

\pacs{Valid PACS appear here}

\maketitle

\section{Introduction}
Freely cooling granular gas under no gravity has been attracting much research interest, because it behaves quite differently from standard gas \cite{Gold2,GG,KT}.
An abstract explanation of its interesting behaviors is as follows. 
The gas system, which is initially prepared in an equilibrium state, cools down uniformly in space \cite{Haff}.
This state is referred to as the homogeneous cooling state (HCS). 
One might naively conjecture that the HCS is stable and that the system will stay in that condition forever.
However, because velocity vectors of colliding particles tend to parallel to each other because of inelastic collisions, a vortex structure develops in the velocity field \cite{TPC}.
This instability is called shearing instability.
Moreover, dense clusters of particles are formed due to what is called clustering instability \cite{Gold}.
After the emergence of clustering instability, clusters of particles collide with each other, merge, and split in a complex way \cite{Luding}. The system eventually develops a high-density region where ``inelastic collapse'' is likely to happen when a constant coefficient of restitution is smaller than a certain critical value \cite{MY1,MY2}.

The theory of granular gas might be useful for understanding the behavior of astrophysical objects such as planetesimals, cosmic dust, and planetary rings.
In general, however, the dynamics of these systems is overwhelmingly controlled by self-gravity, the gravitational force generated by the system itself.
The effect of self-gravitation leads to clustering of particles---the Jeans instability \cite{Jeans,Chandra}.
This instability plays a central important role in the process of formation of astrophysical objects \cite{GalaDy}.
Therefore, in order to apply the theory of granular gas to astrophysical objects, it is necessary to study the dynamics of granular gas that reflects the contribution of self-gravitational force, which therefore exhibits the effects of both inelastic collisions and self-gravitational force.
Hereafter, we call such a system ``self-gravitating granular gas''.

Note that the clustering instability and the Jeans instability are alike in that both of them form dense clusters of particles.
Therefore we expect that, in self-gravitating granular gas, competition between clustering instability and Jeans instability will happen.
If so, it seems natural to ask (i) what characterizes the two instabilities and (ii) what the condition of competition between them is.
In this paper, in order to answer these questions, we investigate the dynamics of self-gravitating granular gas.
Our approach to the problems was as follows.
First, we derived hydrodynamic equations for a gas of hard spheres with dissipative dynamics reflecting the contribution of self-gravitational force of mass density perturbations from the Boltzmann equation.
Here, because it is difficult to incorporate the effect of direct particle-particle interaction in the collision integral of the Boltzmann equation, we neglected the effect of the particle-particle interaction in collisions for simplicity. Instead, we considered a system for which the collision integral is the same as the one that appears in a non-self-gravitating granular gas and in which the effect of self-gravity is incorporated only through mass density perturbations.
This approximation is reasonable considering the fact that the effect of direct particle-particle interaction is much weaker than a many-body gravitational effect.
We then considered the homogeneous solution of the hydrodynamic equations with spatially uniform density.
As is well known, in long-range interacting systems, infinite homogeneous distribution of matter cannot occur.
However, for the purpose of studying linear stability, we removed this difficulty by imposing the so-called ``Jeans swindle'', which is constituted of the ad hoc assumption that Poisson's equation covering the gravitational potential, describes only the relation between the perturbed density and the perturbed potential, while the unperturbed potential is zero \cite{GalaDy}.
Then we analyzed the linear stability of self-gravitating granular gas. 
The results are straightforward: (i) The condition of competition between the clustering instability and the Jeans instability depends on three parameters: the mass density $\rho_0$, the collision rate $\omega_0$, and the rate of energy loss per collision $\epsilon=1-e^2$, where $e$ is a constant coefficient of normal restitution.
When $\sqrt{G\rho_0}\ll\epsilon\omega_0$, clustering instability dominates, while when $\epsilon\omega_0\ll\sqrt{G\rho_0}$, Jeans instability dominates.
(ii) These instabilities are characterized by the decrease and increase, respectively, of the temperature.
In other words, when we increase $\sqrt{G\rho_0}/\epsilon\omega_0$ from small to large, the instability of the system continuously changes from clustering instability to Jeans instability; there is a smooth \textit{crossover} from the clustering instability limit to the Jeans instability limit.

The organization of this paper is as follows.
In Sec. \ref{2}, using dimensional analysis, the mechanisms of clustering instability and Jeans instability is reviewed, and the condition of competition between them is also investigated. 
The condition of competition between the two instabilities are easily obtained in this section. 
In Sec. \ref{3}, starting from the Boltzmann equation that describes self-gravitating granular gas at kinetic scale, we derive the hydrodynamic equations.
Then, the spatially homogeneous solution of the hydrodynamic equations is investigated.
In Sec. \ref{4} we show the results of linear stability analysis.
Discussions and conclusions are presented in Sec. \ref{5}.

\section{Dimensional analysis\label{2}}
Before considering basic equations, we review the mechanisms of clustering instability and Jeans instability.
We also show some results about the competition between the two instabilities using dimensional analysis.

First, we consider clustering instability as follows.
Let us compare $t_s$, the characteristic time scale of the density fluctuation, and $t_c$, the characteristic time scale of the energy dissipation in collisions.
Let $T$ be the temperature of the system, and $m$ the mass of a particle.
Then, as the sound velocity is evaluated as $\sqrt{T/m}$, $t_s$ becomes
\begin{eqnarray}
t_s\sim\dfrac{R}{\sqrt{T/m}},
\label{eq:ts}
\end{eqnarray}
where $R$ is the length scale of the density fluctuation.
Here, we have implicitly assumed that both sides of (\ref{eq:ts}) have the same order of magnitude.
On the other hand, we speculate that $t_c$ depends on the product of the collision rate $\omega_0=\sqrt{T/m}/l_{\rm{mfp}}$ and the rate of energy loss per collision $\epsilon$, where $l_{\rm{mfp}}$ is the mean free path.
Then, using dimensional analysis, we obtain
\begin{eqnarray}
t_c\sim\dfrac{1}{\epsilon \omega_0}=\dfrac{l_{\rm{mfp}}}{\epsilon \sqrt{T/m}}.
\end{eqnarray}
Comparing $t_s$ and $t_c$, we notice that, in the density fluctuation that satisfies $t_c< t_s$, i.e.,
\begin{eqnarray}
\dfrac{l_{\rm{mfp}}}{\epsilon}< R,
\end{eqnarray}
non-uniformity of the density field develops because the effect of the energy loss by inelastic collisions overwhelms the effect of the restoring force that tends to form the spatially uniform density field. This is the mechanism of clustering instability.

Next, we consider the Jeans instability in the same way \cite{GalaDy}.
Let $t_J$ be the characteristic time scale for the system to shrink by its own gravity.
We speculate that $t_J$ depends on the mass density $\rho_0$ and the gravitational constant $G$.
Thus, we obtain
\begin{eqnarray}
t_J\sim\dfrac{1}{\sqrt{G\rho_0}}.
\end{eqnarray}
Comparing $t_s$ and $t_J$, we notice that, in the density fluctuation that satisfies $t_J< t_s$, i.e.,
\begin{eqnarray}
\sqrt{\dfrac{T/m}{G\rho_0}}< R,
\end{eqnarray}
the non-uniformity of the density field develops because the effect of shrinking by self-gravitational force overwhelms the effect of the restoring force. This is the mechanism of Jeans instability.

Finally, we consider the competition between clustering instability and Jeans instability.
Comparing $t_c$ and $t_J$, we notice that when $t_J\ll t_c$, i.e.,
\begin{eqnarray}
\epsilon\omega_0\ll \sqrt{G\rho_0},
\end{eqnarray}
the effect of shrinking by self-gravitation is predominant;
thus, Jeans instability dominates.
At the other limit, $t_c\ll t_J$, i.e.,
\begin{eqnarray}
\sqrt{G\rho_0}\ll\epsilon\omega_0,
\end{eqnarray}
clustering instability dominates.

In summary, we conjecture that the condition of competition between clustering instability and Jeans instability depends on three parameters: $\epsilon$, $\omega_0$, and $\rho_0$.
When $\sqrt{G\rho_0}\ll\epsilon\omega_0$, clustering instability dominates, while when $\epsilon\omega_0\ll\sqrt{G\rho_0}$, Jeans instability dominates.
Below, we study this conjecture more quantitatively by considering hydrodynamic equations.

\section{Hydrodynamic equations\label{3}}
\subsection{Derivation of hydrodynamic equations}
Here, we consider a gas of smooth identical spheres of diameter $\sigma$ and mass $m$, of which the collisions are characterized by a constant coefficient of normal restitution $e$ ($0<e<1$). 
The rotational motion is ignored.
At sufficiently low density, the time evolution of the one-particle distribution function $f(\bm{r},\bm{c},t)$ is governed by the Boltzmann equation \cite{Gold&Shapiro,BDKS}.
Here, for simplicity, we neglect the effect of direct particle-particle interaction in collisions because it is difficult to incorporate the effect of the direct interaction in the collision integral of the Boltzmann equation. 
Instead, we consider a system for which the collision integral is the same as the one that appears in non-self-gravitating granular gas and in which the effect of self-gravitation is incorporated only through mass density perturbations.
This approximation is reasonable considering the fact that the effect of direct particle-particle interaction is much weaker than a many-body gravitational effect.
In other words, we impose the condition that the gas be almost ideal. 
That is, the gravitational energy is far less than the kinetic energy $Gm^2/\bar{r}\ll m\langle\bm{c}^2\rangle/2$, where $m\langle\bm{c}^2\rangle/2$ is the average of the single-particle kinetic energy and $\bar{r}$ is the mean distance between particles. 
Then, the Boltzmann equation for describing self-gravitating granular gas is 
\begin{equation}
\dfrac{\partial f}{\partial t}+\bm{c}\cdot\dfrac{\partial f}{\partial\bm{r}}-\nabla \phi\cdot\dfrac{\partial f}{\partial \bm{c}}=I(f,f),
\label{eq:B_eq}
\end{equation}
where $\phi$ is the gravitational potential, which satisfies Poisson's equation
\begin{eqnarray}
\nabla^2 \phi &=& 4\pi G \rho, \label{eq:Poisson}
\end{eqnarray}
and $I(f,f)$ is the Boltzmann collision integral
\begin{widetext}
\begin{eqnarray}
I(f,f):={\displaystyle \int} d\bm{c}_1{\displaystyle \int}d\hat{\bm{\sigma}}\Theta(-\bm{c}_{\rm{rel}}\cdot\hat{\bm{\sigma}})\sigma^2|\bm{c}_{\rm{rel}}\cdot\hat{\bm{\sigma}}|\left(\dfrac{1}{e^2}f(\bm{r},\bm{c}',t)f(\bm{r},\bm{c}'_1,t)-f(\bm{r},\bm{c},t)f(\bm{r},\bm{c}_1,t)\right).
\end{eqnarray}
\end{widetext}
In the above expression, $\hat{\bm{\sigma}}$ is the unit vector along the line joining the centers of the colliding pair, $\Theta$ is the Heaviside step function, $\bm{c}_{\rm{rel}}=\bm{c}-\bm{c}_1$, and 
\begin{equation}
\bm{c}'=\bm{c} - \dfrac{1+e}{2}(\bm{c}_{\rm{rel}}\cdot\hat{\bm{\sigma}})\hat{\bm{\sigma}},
\end{equation}
\begin{equation}
\bm{c}'_1=\bm{c}_1 + \dfrac{1+e}{2}(\bm{c}_{\rm{rel}}\cdot\hat{\bm{\sigma}})\hat{\bm{\sigma}}, 
\end{equation}
are velocities after a binary collision of particles of which the velocities are $\bm{c}$ and $\bm{c}_1$.

Now, in order to derive the hydrodynamic equations, let us assume scale separation: $l_{\rm{micro}}\ll l_{\rm{macro}}$ and $t_{\rm{micro}}\ll t_{\rm{macro}}$, where $l_{\rm{micro}}$ ($t_{\rm{micro}}$) is the maximum length (time) scale appearing in the microscopic description such as the diameter or the mean free path (mean free time) and $l_{\rm{macro}}$ ($t_{\rm{macro}}$) is the minimum length (time) scale characterizing macroscopic behaviors.
The macroscopic variables of interest are the hydrodynamic fields: the mass density $\rho(\bm{r},t)$, the flow velocity $\bm{v}(\bm{r},t)$, and the granular temperature $T(\bm{r},t)$ are defined in the usual way,
\begin{eqnarray}
\rho(\bm{r},t)={\displaystyle \int} d\bm{c}mf(\bm{r},\bm{c},t), 
\end{eqnarray}
\begin{eqnarray}
\bm{v}(\bm{r},t)=\dfrac{1}{n(\bm{r},t)}{\displaystyle \int} d\bm{c}\bm{c}f(\bm{r},\bm{c},t),
\end{eqnarray}
\begin{equation}
\dfrac{3}{2}n(\bm{r,t})T(\bm{r},t)={\displaystyle \int} d\bm{c}\dfrac{1}{2}m(\bm{c}-\bm{v}(\bm{r},t))^2f(\bm{r},\bm{c},t),
\end{equation} 
where $n(\bm{r},t)$ is the number density defined by $\rho(\bm{r},t)/m$.
Here, we assume that $\epsilon\ll 1$ in order for the time scale $t_{\rm{temperature}}$, which characterizes the behavior of the temperature, to satisfy $t_{\rm{micro}}\ll t_{\rm{temperature}}$.

The macroscopic balance equations for these variables are obtained from the following properties of the collision integral
\begin{eqnarray}
{\displaystyle \int}d\bm{c}\left(
\begin{array}{c}
1\\
\bm{c}\\
\dfrac{1}{2}m(\bm{c}-\bm{v})^2
\end{array}
\right)I(f,f)=\left(
\begin{array}{c}
0\\
\bm{0}\\
-\gamma
\end{array}
\right).
\end{eqnarray}
The first two equations follow from the conservation of mass and momentum in collisions.
The last equation reflects the loss of energy in inelastic collisions, and $\gamma$ is the energy dissipation rate defined by
\begin{eqnarray}
\gamma=-{\displaystyle \int} d\bm{c}\dfrac{1}{2}m\bm{c}^2I(f,f).
\end{eqnarray}
From the above properties and (\ref{eq:B_eq}), the following evolution equations for the hydrodynamic fields are easily obtained:
\begin{equation} 
\dfrac{\partial\rho}{\partial t}= -\nabla\cdot(\rho\bm{v}), \label{eq:basic_1}
\end{equation}
\begin{equation}
\rho\left(\dfrac{\partial\bm{v}}{\partial t}+\bm{v}\cdot\nabla\bm{v}\right) =-\nabla \cdot \mathsf{P}-\rho \nabla \phi, \label{eq:basic_2}
\end{equation}
\begin{equation}
\dfrac{3}{2}n \left(\dfrac{\partial T}{\partial t}+\bm{v}\cdot\nabla T\right) =-\nabla \cdot \bm{Q}-\mathsf{P}:\mathsf{D}-\gamma, \label{eq:basic_3}
\end{equation}
\begin{equation}
\nabla^2 \phi = 4\pi G \rho, \label{eq:basic_4}
\end{equation}
where
\begin{eqnarray}
\mathsf{D}_{ij}=\dfrac{1}{2}(\partial_iv_j+\partial_jv_i) 
\end{eqnarray}
is the strain rate tensor,
\begin{multline}
\mathsf{P}_{ij}=n T\delta_{ij} \\
+{\displaystyle \int} d\bm{c}m\left((c_i-v_i)(c_j-v_j)-\dfrac{1}{3}\delta_{ij}(\bm{c}-\bm{v})^2\right)f(\bm{r},\bm{c},t)
\end{multline}
is the stress tensor,
\begin{eqnarray}
\bm{Q}={\displaystyle \int} d\bm{c}\dfrac{1}{2}m(\bm{c}-\bm{v})^2(\bm{c}-\bm{v})f(\bm{r},\bm{c},t)
\end{eqnarray}
is the heat flux.
To reduce (\ref{eq:basic_2}) and (\ref{eq:basic_3}) to the standard equations of fluid mechanics, however, we still have to express $\mathsf{P}$, $\bm{Q}$, and $\gamma$ in terms of macroscopic quantities.
This is achieved by Grad's method \cite{JR,Grad} and the Chapman-Enskog method \cite{CC,BDKS,KT}.
In Grad's method, the distribution function is expanded in a complete set of orthogonal polynomials appropriate to the problem.
The Chapman-Enskog method assumes the existence of a solution of which the space and time dependence is given entirely through the hydrodynamic variables and their gradients.
As a result, $\mathsf{P}$, $\bm{Q}$, and $\gamma$ are given in terms of these variables, and (\ref{eq:basic_1})-(\ref{eq:basic_4}) become a closed set of hydrodynamic equations.
For the leading order in the spatial gradients, the stress tensor and the heat flux are found to be given by
\begin{eqnarray}
\mathsf{P}_{ij}=n T\delta_{ij}-2\eta\left(\mathsf{D}_{ij}-\dfrac{1}{3}\mathsf{D}_{kk}\delta_{ij}\right),
\label{eq:constitutive_1}
\end{eqnarray} 
\begin{eqnarray}
\bm{Q}= -\kappa \nabla T,
\label{eq:constitutive_2}
\end{eqnarray}
where $\eta$ is the shear viscosity and $\kappa$ is the thermal conductivity.
The transport coefficients $\eta$, $\kappa$, and $\gamma$ are calculated as follows \cite{Gold}:
\begin{equation}
\eta=\dfrac{5\sqrt{\pi}}{48}\sigma\rho_s\sqrt{T/m}, 
\end{equation}
\begin{equation}
\kappa=\dfrac{25\sqrt{\pi}}{128}\sigma\rho_s\sqrt{T/m},
\end{equation}
\begin{equation}
\gamma= 24\epsilon\dfrac{\nu^2\rho_s}{\sqrt{\pi}\sigma}(T/m)^{3/2},
\end{equation}
where $\rho_s$ is the mass density of a solid particle defined by $\rho_s=3m/4\pi(\sigma/2)^3$, and $\nu$ is the volume fraction defined by $\nu=\rho_0/\rho_s$, where $\rho_0$ is the spatially homogeneous mass density.

\subsection{Homogeneous solution}
Before going to the stability analysis, it is useful to explain the spatially homogeneous solution.
Hereafter, we assume that the system is infinitely extended.
The homogeneous solution of (\ref{eq:basic_1}), (\ref{eq:basic_2}), and (\ref{eq:basic_4}) is given by $\rho=\rho_0\equiv mn_0=\rm{const}$, $\bm{v}=\bm{0}$, and $\phi=\phi_0=\rm{const}$.
The temperature satisfies the equation
\begin{eqnarray}
\dfrac{3}{2}n_0 \dfrac{d T_0(t)}{dt}&=&-\gamma \notag\\
&=&-24\epsilon \dfrac{\nu^2\rho_s}{\sqrt{\pi}\sigma}(T_0/m)^{3/2},
\label{eq:HCSeq}
\end{eqnarray}
where $T_0(t)$ denotes the spatially homogeneous temperature.
The solution of (\ref{eq:HCSeq}) is
\begin{eqnarray}
T_0(t)&=&\dfrac{T_0(0)}{(1+t/t_0)^2},
\end{eqnarray}
where
\begin{eqnarray}
t_0:=\dfrac{\sqrt{\pi}\sigma}{8\nu \epsilon \sqrt{T_0(0)/m}}=\dfrac{l_0}{\epsilon \sqrt{T_0(0)/m}},
\end{eqnarray}
and 
\begin{eqnarray}
l_0 :=\dfrac{\sqrt{\pi}\sigma}{8\nu},
\end{eqnarray}
which is ``the effective mean free path'' \cite{Gold}.
Thus, in the spatially uniform solution, the temperature decreases monotonically in time.
The solution represents the state that is called the homogeneous cooling state (HCS).
For later convenience, we introduce the mean number of collisions $s$ defined by
\begin{eqnarray}
s := \dfrac{1}{\epsilon}\log(1+\dfrac{t}{t_0}).
\end{eqnarray}
Using this, (\ref{eq:HCSeq}) can be rewritten as
\begin{eqnarray}
\dfrac{d T_0(s)}{ds}=-2\epsilon T_0(s).
\end{eqnarray}
Here, we use the same symbol $T_0(\cdot)$.
Thus, the solution of the equation is expressed as
\begin{eqnarray}
T_0(s)=T_0(0)\mathrm{exp}(-2\epsilon s).
\label{eq:HCS_Ts}
\end{eqnarray}

Note that there is difficulty in defining the spatially homogeneous state for self-gravitating systems: we find that $\rho_0=0$ from (\ref{eq:basic_4}).
However, for the purpose of studying linear stability, we remove this difficulty by imposing the so-called ``Jeans swindle'', which is constructed of the ad hoc assumption that Poisson's equation (\ref{eq:basic_4}) describes only the relation between the perturbed density and the perturbed potential, while the unperturbed potential is zero \cite{GalaDy}.

\section{Linear stability analysis\label{4}}
Consider a small perturbation from the HCS by letting
\begin{eqnarray}
\bm{v}&=&\bm{0}+\bm{v}^{(1)}(\bm{r},t), \\
T&=&T_0(t)+T^{(1)}(\bm{r},t), \\
\rho&=&\rho_0+\rho^{(1)}(\bm{r},t), \\
\phi&=&\phi_0+\phi^{(1)}(\bm{r},t),
\end{eqnarray}
where all the variables with subscript 1 represent perturbations.
A set of Fourier transformed dimensionless variables is defined by
\begin{equation}
\Tilde{\bm{u}}(\bm{k},t)=\dfrac{\Tilde{\bm{v}}^{(1)}(\bm{k},t)}{\sqrt{T_0(t)}}={\displaystyle \int}d\bm{r}e^{-i\bm{k}\cdot\bm{r}}\dfrac{\bm{v}^{(1)}(\bm{r},t)}{\sqrt{T_0(t)}},
\end{equation}
\begin{equation}
\Tilde{\theta}(\bm{k},t)=\dfrac{\Tilde{T}^{(1)}(\bm{k},t)}{T_0(t)}={\displaystyle \int}d\bm{r}e^{-i\bm{k}\cdot\bm{r}}\dfrac{T^{(1)}(\bm{r},t)}{T_0(t)},
\end{equation}
\begin{equation}
\Tilde{\omega}(\bm{k},t)=\dfrac{\Tilde{\rho}^{(1)}(\bm{k},t)}{\rho_0}={\displaystyle \int}d\bm{r}e^{-i\bm{k}\cdot\bm{r}}\dfrac{\rho^{(1)}(\bm{r},t)}{\rho_0}. 
\end{equation}
In terms of these variables, the linearization of hydrodynamic equations (\ref{eq:basic_1})-(\ref{eq:basic_4}) around the HCS yields
\begin{equation}
\dfrac{\partial \Tilde{\bm{u}}_{\perp}}{\partial s}=\left(\epsilon - \frac{5}{6}K^2\right)\Tilde{\bm{u}}_{\perp}, \label{eq:linear_1}
\end{equation}
\begin{equation}
\dfrac{\partial \Tilde{u}_{\parallel}}{\partial s}=-iK\Tilde{\theta}-iK\Tilde{\omega}+\left(\epsilon-\frac{10}{9}K^2\right)\Tilde{u}_{\parallel}+i\frac{B}{K}\mathrm{e}^{2\epsilon s}\Tilde{\omega}, \label{eq:linear_2}
\end{equation}
\begin{equation}
\dfrac{\partial \Tilde{\theta}}{\partial s}=-2\epsilon \Tilde{\omega}-\left(\frac{25}{24}K^2+\epsilon\right)\Tilde{\theta}-\frac{2}{3}iK\Tilde{u}_{\parallel}, \label{eq:linear_3}
\end{equation}
\begin{equation}
\dfrac{\partial \Tilde{\omega}}{\partial s}=-iK\Tilde{u}_{\parallel}.\label{eq:linear_4}
\end{equation}
Here, we have eliminated $\phi$ by using Poisson's equation (\ref{eq:basic_4}) and have changed the variable from $t$ to $s$.
The variables $\Tilde{\bm{u}}_{\perp}$ and $\Tilde{u}_{\parallel}$ denote the longitudinal and transverse components, respectively, of the velocity field relative to the wave vector $\bm{k}$.
Moreover, we have introduced the dimensionless wavenumber $K$ defined by
\begin{eqnarray}
K=kl_0,
\end{eqnarray}
and the dimensionless parameter $B$ defined by
\begin{eqnarray}
B=4\pi\dfrac{G\rho_0}{\omega^2_0},
\end{eqnarray}
where $\omega_0$ is the collision rate at HCS, which is given by
\begin{eqnarray}
\omega_0=\dfrac{\sqrt{T_0(0)/m}}{l_0}.
\end{eqnarray}
From the dimensional analysis, we conjecture that the condition of competition between clustering instability and Jeans instability depends on $B$ and $\epsilon$.

The equation (\ref{eq:linear_1}) is decoupled from the other and can be directly integrated yielding
\begin{eqnarray}
\Tilde{\bm{u}}_{\perp}(\bm{k},s)=\Tilde{\bm{u}}_{\perp}(\bm{k},0) \exp\left[\left(\epsilon - \frac{5}{6}K^2\right)s\right].
\end{eqnarray}
This solution represents shear modes.
The growth rate of this perturbation is $\lambda_{\rm{shear}}:=\epsilon - 5K^2/6$.

In what follows, we consider (\ref{eq:linear_2})-(\ref{eq:linear_4}).
In (\ref{eq:linear_2}), note that $\exp(2\epsilon s)\simeq 1$ for finite $s$, because $\epsilon\ll 1$.
Then we seek the solutions of the form
\begin{eqnarray}
\Tilde{u}_{\parallel}(\bm{k},s)&=&\Tilde{u}_{\parallel}(\bm{k},0)\exp(\lambda(K,\epsilon,B)s), \label{eq:normal_mode_1}\\[5pt]
\Tilde{\theta}(\bm{k},s)&=&\Tilde{\theta}(\bm{k},0)\exp(\lambda(K,\epsilon,B)s), \label{eq:normal_mode_2}\\[5pt] 
\Tilde{\omega}(\bm{k},s)&=&\Tilde{\omega}(\bm{k},0)\exp(\lambda(K,\epsilon,B)s), \label{eq:normal_mode_3}
\end{eqnarray}
where $\lambda(K,\epsilon,B)$ denotes the complex growth rate.
A perturbation is unstable when the real part of $\lambda(K,\epsilon,B)$ is positive.
Substituting (\ref{eq:normal_mode_1})-(\ref{eq:normal_mode_3}) into (\ref{eq:linear_2})-(\ref{eq:linear_4}) yields
\begin{multline}
\lambda\left(\lambda-\epsilon+\dfrac{10}{9}K^2\right)\left(\lambda+\epsilon+\dfrac{25}{24}K^2\right)-2\epsilon K^2 \\
+\dfrac{2}{3}K^2\lambda + \left(\lambda+\dfrac{25}{24}K^2+\epsilon\right)(K^2-B)=0.
\label{eq:dispersion}
\end{multline}
Let us denote the three solutions of (\ref{eq:dispersion}) by $\lambda_\alpha(K,\epsilon,B)$ ($\alpha=1,2,3$).

Since we are interested in the small wavenumber range $K\ll 1$, which leads to the unstable fluctuation, we analyze the response of the system using asymptotic expansions: we assume that $\lambda_\alpha(K,\epsilon,B)$ has asymptotic expansions in terms of asymptotic sequences $\{K^n\}^{\infty}_{n=0}$ as $K\rightarrow0$:
\begin{multline}
\lambda_\alpha(K,\epsilon,B)\sim\lambda^{(0)}_\alpha(\epsilon,B)+K\lambda^{(1)}_\alpha(\epsilon,B) \\
+K^2\lambda^{(2)}_\alpha(\epsilon,B)+\cdots.
\label{eq:asymptotic_expansion}
\end{multline}
Substituting this expansion into (\ref{eq:dispersion}), we can calculate $\lambda^{(1)}_\alpha, \lambda^{(2)}_\alpha, \cdots$ (see Appendix).
One of the three solutions of the above equation, $\lambda_1$, can be positive and corresponds to the unstable mode (heat mode), the other two solutions, $\lambda_{2,3}$, are a complex conjugate pair and correspond to sound modes. 
From now on, we concentrate our attention on $\lambda_1$.
Substituting the asymptotic expansion of $\lambda_1$ up to the second order into (\ref{eq:normal_mode_1})-(\ref{eq:normal_mode_3}), one obtains
\begin{widetext}
\begin{equation}
\Tilde{\rho}^{(1)}(\bm{k},s)\simeq\Tilde{\rho}^{(1)}(\bm{k},0)\exp\left[\left(\dfrac{\epsilon}{2}+\sqrt{\dfrac{\epsilon^2}{4}+B}+\dfrac{-\dfrac{10}{9}(\epsilon^2+B)-\left(\dfrac{20}{9}\epsilon+\dfrac{5}{3}\right)\sqrt{\dfrac{\epsilon^2}{4}+B}+\dfrac{\epsilon}{6}}{\dfrac{\epsilon^2}{2}+2B+3\epsilon\sqrt{\dfrac{\epsilon^2}{4}+B}}K^2\right)s\right], 
\end{equation}
\begin{equation}
\Tilde{v}^{(1)}_{\parallel}(\bm{k},s)\simeq\Tilde{v}^{(1)}_{\parallel}(\bm{k},0)\exp\left[\left(-\dfrac{\epsilon}{2}+\sqrt{\dfrac{\epsilon^2}{4}+B}
+\dfrac{-\dfrac{10}{9}(\epsilon^2+B)-\left(\dfrac{20}{9}\epsilon+\dfrac{5}{3}\right)\sqrt{\dfrac{\epsilon^2}{4}+B}+\dfrac{\epsilon}{6}}{\dfrac{\epsilon^2}{2}+2B+3\epsilon\sqrt{\dfrac{\epsilon^2}{4}+B}}K^2\right)s\right], 
\end{equation}
\begin{equation}
\Tilde{T}^{(1)}(\bm{k},s)\simeq\Tilde{T}^{(1)}(\bm{k},0)\exp\left[\left(-\dfrac{3}{2}\epsilon+\sqrt{\dfrac{\epsilon^2}{4}+B}
+\dfrac{-\dfrac{10}{9}(\epsilon^2+B)-\left(\dfrac{20}{9}\epsilon+\dfrac{5}{3}\right)\sqrt{\dfrac{\epsilon^2}{4}+B}+\dfrac{\epsilon}{6}}{\dfrac{\epsilon^2}{2}+2B+3\epsilon\sqrt{\dfrac{\epsilon^2}{4}+B}}K^2\right)s\right].
\end{equation}
\end{widetext}
Here, we have changed the variables from $\Tilde{u}_{\parallel}$, $\Tilde{\theta}$, $\Tilde{\omega}$ to $\Tilde{v}^{(1)}_{\parallel}$, $\Tilde{T}^{(1)}$, $\Tilde{\rho}^{(1)}$.
Since the relationship between $B$ and $\epsilon$ determines the competition between clustering instability and Jeans instability, we consider the limits of both $B\ll \epsilon^2$ ($\epsilon\omega_0\ll\sqrt{G\rho_0}$) and $\epsilon^2 \ll B$ ($\sqrt{G\rho_0}\ll\epsilon\omega_0$), as we have seen in Sec. \ref{2}.

When $B\ll\epsilon^2$, we conjecture that the clustering instability dominates.
To verify this conjecture, let us check the sign of coefficients in front of $s$ by approximating them using $B\ll\epsilon^2$:
\begin{widetext}
\begin{equation}
\Tilde{\rho}^{(1)}(\bm{k},s)\simeq\Tilde{\rho}^{(1)}(\bm{k},0)\exp\left(\left\{\dfrac{\epsilon}{2}+\sqrt{\dfrac{\epsilon^2}{4}+B}+\left[-\dfrac{10}{9}\left(1+\dfrac{3}{10\epsilon}\right)+O\left(\dfrac{B}{\epsilon^2}\right)\right]K^2\right\}s\right), 
\end{equation}
\begin{equation}
\Tilde{v}^{(1)}_{\parallel}(\bm{k},s)\simeq\Tilde{v}^{(1)}_{\parallel}(\bm{k},0)\exp\left(\left\{\dfrac{B}{\epsilon}+O\left(\left(\dfrac{B}{\epsilon^2}\right)^2\right)+\left[-\dfrac{10}{9}\left(1+\dfrac{3}{10\epsilon}\right)+O\left(\dfrac{B}{\epsilon^2}\right)\right]K^2\right\}s\right),
\end{equation}
\begin{equation}
\Tilde{T}^{(1)}(\bm{k},s)\simeq\Tilde{T}^{(1)}(\bm{k},0)\exp\left(\left\{-\epsilon\left(1+O\left(\dfrac{B}{\epsilon^2}\right)\right)+\left[-\dfrac{10}{9}\left(1+\dfrac{3}{10\epsilon}\right)+O\left(\dfrac{B}{\epsilon^2}\right)\right]K^2\right\}s\right).
\end{equation}
\end{widetext}
The above result shows that the density fluctuation is unstable and that the temperature fluctuation is stable.
This is consistent with the fact that, in the clustering instability, the kinetic energy decreases due to inelastic collisions.

When $\epsilon^2\ll B$, we conjecture that the Jeans instability dominates.
To verify this conjecture, let us check the sign of coefficients in front of $s$ by approximating them using $\epsilon^2\ll B$:
\begin{widetext}
\begin{equation}
\Tilde{\rho}^{(1)}(\bm{k},s)\simeq\Tilde{\rho}^{(1)}(\bm{k},0)\exp\left(\left\{\dfrac{\epsilon}{2}+\sqrt{\dfrac{\epsilon^2}{4}+B}+\left[-\dfrac{5}{9}\left(1+\dfrac{3}{2\sqrt{B}}\right)+O\left(\left(\dfrac{\epsilon^2}{B}\right)^{1/2}\right)\right]K^2\right\}s\right), 
\end{equation}
\begin{equation}
\Tilde{v}^{(1)}_{\parallel}(\bm{k},s)\simeq\Tilde{v}^{(1)}_{\parallel}(\bm{k},0)\exp\left(\left\{\sqrt{B}\left[1+O\left(\left(\dfrac{\epsilon^2}{B}\right)^{1/2}\right)\right]+\left[-\dfrac{5}{9}\left(1+\dfrac{3}{2\sqrt{B}}\right)+O\left(\left(\dfrac{\epsilon^2}{B}\right)^{1/2}\right)\right]K^2\right\}s\right), 
\end{equation}
\begin{equation}
\Tilde{T}^{(1)}(\bm{k},s)\simeq\Tilde{T}^{(1)}(\bm{k},0)\exp\left(\left\{\sqrt{B}\left[1+O\left(\left(\dfrac{\epsilon^2}{B}\right)^{1/2}\right)\right]+\left[-\dfrac{5}{9}\left(1+\dfrac{3}{2\sqrt{B}}\right)+O\left(\left(\dfrac{\epsilon^2}{B}\right)^{1/2}\right)\right]K^2\right\}s\right).
\end{equation}
\end{widetext}
Note that, different from the clustering instability, the temperature fluctuation is unstable as well as the density fluctuation.
This result is consistent with the fact that, in the Jeans instability, the gravitational potential energy transforms into kinetic energy.

\begin{figure}[h]
 \includegraphics[width=8.6cm]{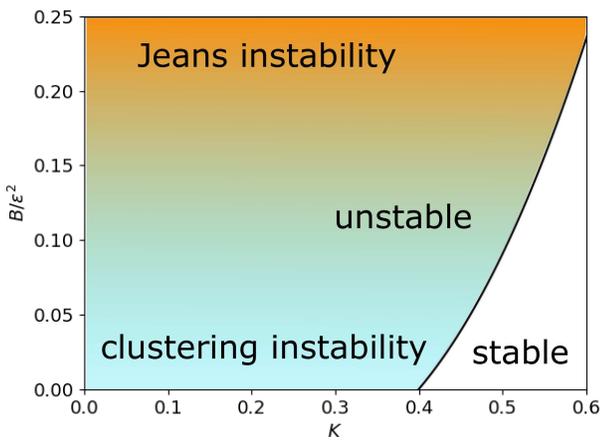}
 \caption{(Color online) Phase diagram for ($K$, $B/\epsilon^2$). The solid line corresponds to the graph of (\ref{eq:critical wavenumber}). Here, $\epsilon$ is fixed as $0.1$.}
 \label{fig:crossover}
\end{figure}

Furthermore, substituting $\lambda=0$ into (\ref{eq:dispersion}), one obtains the critical value of the wavelength
\begin{equation}
K^*=\sqrt{\dfrac{12}{25}\left(\epsilon+\dfrac{25}{24}B+\sqrt{\left(\epsilon+\dfrac{25}{24}B\right)^2+\dfrac{25}{6}\epsilon B}\right)}.
\label{eq:critical wavenumber}
\end{equation}
When $B\ll\epsilon^2$, $K^*$ corresponds to the critical value of the wavelength of the clustering instability $k^*_{\rm{cl}}\sim\sqrt{24\epsilon/25}/l_0$ \cite{Gold}.
At the other limit $\epsilon^2\ll B$, $K^*$ corresponds to that of the Jeans instability $k^*_{J}\sim\sqrt{4\pi Gm\rho_0/T_0(0)}$ \cite{Chandra}.
The solid line in Fig. \ref{fig:crossover} shows $K^*$ dependence of $B/\epsilon^2$.
The perturbation of which the parameters are in the right domain is stable, while the perturbation of which the parameters are in the left domain is unstable.
When we increase $B/\epsilon^2$ from small to large, the instability of the system continuously changes from clustering instability to the Jeans instability; there is a smooth \textit{crossover} from the clustering instability limit to the Jeans instability limit. 
\begin{figure}[t]
\includegraphics[width=8.6cm]{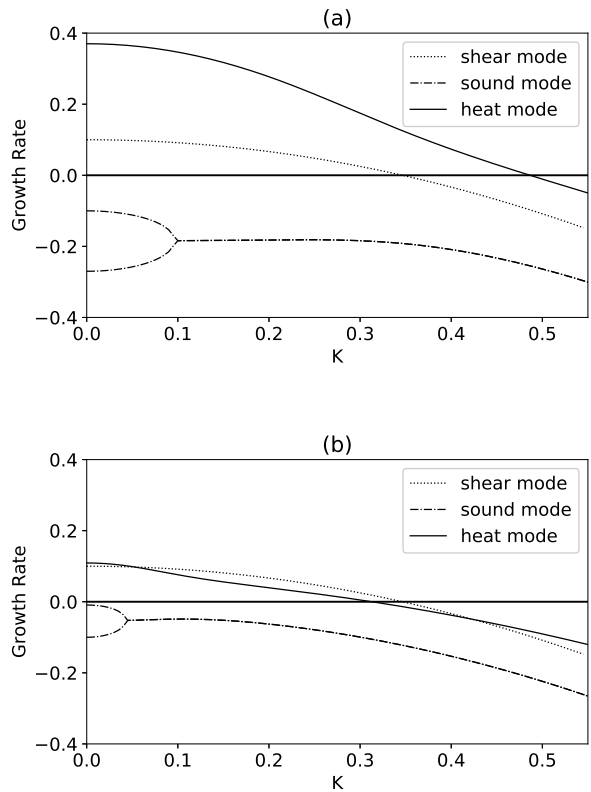}
\caption{Growth rate of disturbances versus reduced wavenumber $K$ for the hydrodynamic modes. The chosen parameter values are ($\epsilon$, $B$)=($0.1$, $0.1$) in (a), and ($0.1$, $0.001$) in (b).}
\label{fig:e=01_B=01_B=0001}
\end{figure}

It is straightforward to numerically solve (\ref{eq:dispersion}) in the entire parameter space.
The $K$ dependence of the real parts of $\lambda_{1,2,3}$ and $\lambda_{\rm{shear}}$ for $\epsilon=0.1$ and $B=0.1,0.001$ are shown in Fig. \ref{fig:e=01_B=01_B=0001}(a) and Fig. \ref{fig:e=01_B=01_B=0001}(b), respectively.
The solid line corresponds to the unstable mode $\lambda_1$, the dashed line to the shear modes $\lambda_{\rm{shear}}$, and the dot-dash line to the sound modes $\lambda_{2,3}$.
These graphs show that the growth rate of the Jeans instability is larger than that of the clustering instability.
In addition, Fig. \ref{fig:e=01_B=01_B=0001}(a) and Fig. \ref{fig:e=01_B=01_B=0001}(b) show that the growth rate of the shearing instability is larger than that of the clustering instability, while it is smaller than that of the Jeans instability.

\section{Concluding remarks \label{5}}
In summary, we investigated the linear stability of self-gravitating granular gas that exhibits the effect of both inelastic collision and self-gravitational force.
Using linear stability analysis and dimensional analysis, we clarified the condition of competition between clustering instability and Jeans instability: when $\sqrt{G\rho_0}\ll\epsilon\omega_0$, the clustering instability dominates, and vice versa.
We also revealed that the two instabilities are characterized by the decrease and increase, respectively, of the temperature.
In other words, when we continuously increase the parameters from $\sqrt{G\rho_0}\ll\epsilon\omega_0$ to $\epsilon\omega_0\ll\sqrt{G\rho_0}$, the system undergoes a smooth crossover from the clustering instability limit to the Jeans instability limit.

The linear stability analysis presented in this paper describes merely the onset of the instability.
The actual structure-forming process that takes place after onset of the instability is governed by the nonlinearities of the hydrodynamic equations.
Moreover, we investigated merely a simple model in which the direct particle-particle interaction in collisions is neglected and of which the ranges of parameters are limited in order for the hydrodynamic approximation to be valid. 
Thus, the task of investigating the growth of the perturbations in the nonlinear regime and considering the effect of more realistic direct particle-particle interaction in collisions should be addressed in the future.

\begin{acknowledgments}
The author thanks Shin-ichi Sasa, Minoru Sekiya, and Hiizu Nakanishi for their fruitful comments. 
\end{acknowledgments}

\appendix*
\section{}
Substituting the asymptotic expansion (\ref{eq:asymptotic_expansion}) up to the second order into (\ref{eq:dispersion}) gives
\begin{widetext}
\begin{multline}
\lambda^{(0)}(\lambda^{(0)}-\epsilon)(\lambda^{(0)}+\epsilon)-B(\lambda^{(0)}+\epsilon)+K\left[\lambda^{(1)}(\lambda^{(0)}-\epsilon)(\lambda^{(0)}+\epsilon)+2\lambda^{(1)}(\lambda^{(0)})^2-B\lambda^{(1)}\right] \\
+K^2\left\{\lambda^{(2)}(\lambda^{(0)}-\epsilon)(\lambda^{(0)}+\epsilon)+2(\lambda^{(1)})^2\lambda^{(0)}+\lambda^{(0)}\left[(\lambda^{(0)}+\epsilon)\left(\lambda^{(2)}+\dfrac{10}{9}\right)+(\lambda^{(0)}-\epsilon)\left(\lambda^{(2)}+\dfrac{25}{24}\right)+(\lambda^{(1)})^2\right] \right.\\
\left.-\epsilon+\dfrac{5}{3}\lambda^{(0)}-B\left(\lambda^{(2)}+\dfrac{25}{24}\right)\right\}+O(K^3)=0.
\end{multline}
\end{widetext}
Setting $K=0$, one gets a cubic equation for $\lambda^{(0)}_{1,2,3}$:
\begin{eqnarray}
\lambda^{(0)}(\lambda^{(0)}-\epsilon)(\lambda^{(0)}+\epsilon)-B(\lambda^{(0)}+\epsilon)=0,
\end{eqnarray}
and can determine $\lambda^{(0)}_{1,2,3}$.
Repeating this procedure, one can determine $\lambda^{(1)}_{1,2,3}$, $\lambda^{(2)}_{1,2,3}$, and so on.
Thus, we finally get the asymptotic expansion of $\lambda_{1,2,3}$ up to the second order:
\begin{widetext}
\begin{eqnarray}
\lambda_1&\sim&\dfrac{\epsilon}{2}+\sqrt{\dfrac{\epsilon^2}{4}+B}+\dfrac{-\dfrac{10}{9}(\epsilon^2+B)-\left(\dfrac{20}{9}\epsilon+\dfrac{5}{3}\right)\sqrt{\dfrac{\epsilon^2}{4}+B}+\dfrac{\epsilon}{6}}{\dfrac{\epsilon^2}{2}+2B+3\epsilon\sqrt{\dfrac{\epsilon^2}{4}+B}}K^2+O(K^3), \label{eq:lambda21}\\[6pt]
\lambda_2&\sim&\dfrac{\epsilon}{2}-\sqrt{\dfrac{\epsilon^2}{4}+B}+\dfrac{-\dfrac{10}{9}(\epsilon^2+B)+\left(\dfrac{20}{9}\epsilon+\dfrac{5}{3}\right)\sqrt{\dfrac{\epsilon^2}{4}+B}+\dfrac{\epsilon}{6}}{\dfrac{\epsilon^2}{2}+2B-3\epsilon\sqrt{\dfrac{\epsilon^2}{4}+B}}K^2+O(K^3), \label{eq:lambda22}\\[6pt]
\lambda_3&\sim&-\epsilon+\dfrac{-\dfrac{25}{12}\epsilon^2+\dfrac{8}{3}\epsilon+\dfrac{25}{24}B}{2\epsilon^2-B}K^2+O(K^3). \label{eq:lambda23}
\label{eq:mode}
\end{eqnarray}
\end{widetext}
At quite small $K$ all modes are real, while at larger $K$ two modes, $\lambda_2$ and $\lambda_3$, become a complex conjugate pair of propagating modes.

\end{document}